# A giant impact as the likely origin of different twins in the Kepler-107 exoplanet system

Aldo S. Bonomo[1*], Li Zeng[2], Mario Damasso[1], Zoë M. Leinhardt[3], Anders B. Justesen[4], Eric Lopez[5], Mikkel N. Lund[4], Luca Malavolta[6,7], Victor Silva Aguirre[4], Lars A. Buchhave[8], Enrico Corsaro[9], Thomas Denman[3], Mercedes Lopez-Morales[10], Sean M. Mills[11], Annelies Mortier[12], Ken Rice[13], Alessandro Sozzetti[1], Andrew Vanderburg[10,14], Laura Affer[15], Torben Arentoft[4], Mansour Benbakoura[16,17], François Bouchy[18], Jørgen Christensen-Dalsgaard[4], Andrew Collier Cameron[12], Rosario Cosentino[19], Courtney D. Dressing[20], Xavier Dumusque[18], Pedro Figueira[21,22], Aldo F. M. Fiorenzano[19], Rafael A. García[16,17], Rasmus Handberg[4], Avet Harutyunyan[19], John A. Johnson[10], Hans Kjeldsen[4], David W. Latham[10], Christophe Lovis[18], Mia S. Lundkvist[4,23], Savita Mathur[24,25], Michel Mayor[18], Giusi Micela[15], Emilio Molinari[26], Fatemeh Motalebi[18], Valerio Nascimbeni[6,7], Chantanelle Nava[10], Francesco Pepe[18], David F. Phillips[10], Giampaolo Piotto[6,7], Ennio Poretti[19,27], Dimitar Sasselov[10], Damien Ségransan[18], Stéphane Udry[18] & Chris Watson[28]

---

[1]INAF–Osservatorio Astrofisico di Torino, via Osservatorio 20, 10025 Pino Torinese, Italy; [2]Department of Earth and Planetary Sciences, Harvard University, Cambridge, MA 02138, USA; [3]School of Physics, University of Bristol, HH Wills Physics Laboratory, Tyndall Avenue, Bristol BS8 1TL, UK; [4]Stellar Astrophysics Centre, Department of Physics and Astronomy, Aarhus University, Ny Munkegade 120, DK-8000 Aarhus C, Denmark; [5]NASA Goddard Space Flight Center, 8800 Greenbelt Road, Greenbelt, MD 20771, USA; [6]Dipartimento di Fisica e Astronomia "Galileo Galilei", Università di Padova, Vicolo dell'Osservatorio 3, I-35122 Padova, Italy; [7]INAF—Osservatorio Astronomico di Padova, Vicolo dell'Osservatorio 5, I-35122 Padova, Italy; [8]DTU Space, National Space Institute, Technical University of Denmark, Elektrovej 328, DK-2800 Kgs. Lyngby; [9]INAF - Osservatorio Astrofisico di Catania, via S. Sofia 78, 95123, Catania, Italy; [10]Harvard-Smithsonian Center for Astrophysics, 60 Garden Street, Cambridge, MA 02138, USA; [11]The Department of Astronomy, The California Institute of Technology, 1200 East California Blvd, Pasadena, CA 91125, USA; [12]Centre for Exoplanet Science, SUPA, School of Physics and Astronomy, University of St Andrews, St Andrews KY16 9SS, UK; [13]SUPA, Institute for Astronomy, University of Edinburgh, Royal Observatory, Blackford Hill, Edinburgh, EH93HJ, UK; [14]Department of Astronomy, The University of Texas at Austin, 2515 Speedway, Stop C1400, Austin, TX 78712, USA; [15]INAF—Osservatorio Astronomico di Palermo, Piazza del Parlamento 1, I-90124 Palermo, Italy; [16]IRFU, CEA, Université Paris-Saclay, F-91191 Gif-sur-Yvette, France; [17]Université Paris Diderot, AIM, Sorbonne Paris Cité, CEA, CNRS, F-91191 Gif-sur-Yvette, France; [18]Observatoire Astronomique de l'Université de Genève, 51 ch. des Maillettes, 1290 Versoix, Switzerland; [19]INAF—Fundación Galileo Galilei, Rambla José Ana Fernandez Pérez 7, E-38712 Breña Baja, Spain; [20]Astronomy Department, University of California Berkeley, Berkeley, CA 94720-3411, USA; [21]European Southern Observatory, Alonso de Cordova 3107, Vitacura, Region Metropolitana, Chile; [22]Instituto de Astrofísica e Ciências do Espaço, Universidade do Porto, CAUP, Rua das Estrelas, PT4150-762 Porto, Portugal; [23]Zentrum für Astronomie der Universität Heidelberg, Landessternwarte, Königstuhl 12, 69117 Heidelberg, Germany; [24]Departamento de Astrofísica, Universidad de La Laguna, E-38206, Tenerife, Spain; [25]Instituto de Astrofísica de Canarias, C/ Vía Láctea s/n, E-38205, La Laguna, Tenerife, Spain; [26]INAF—Osservatorio Astronomico di Cagliari, Via della Scienza 5—I-09047 Selargius (CA), Italy; [27]INAF – Osservatorio Astronomico di Brera, via E. Bianchi 46, 23807 Merate (LC), Italy; [28]Astrophysics Research Centre, School of Mathematics and Physics, Queen's University Belfast, Belfast BT7 1NN, UK. *e-mail: aldo.bonomo@inaf.it

**Measures of exoplanet bulk densities indicate that small exoplanets with radius less than 3 Earth radii ( $R_\oplus$ ) range from low-density sub-Neptunes containing volatile elements[1] to higher density rocky planets with Earth-like[2] or iron-rich[3] (Mercury-like) compositions. Such astonishing diversity in observed small exoplanet compositions may be the product of different initial conditions of the planet-formation process and/or different evolutionary paths that altered the planetary properties after formation[4]. Planet evolution may be especially affected by either photoevaporative mass loss induced by high stellar X-ray and extreme ultraviolet (XUV) flux[5] or giant impacts[6]. Although there is some evidence for the former[7,8], there are no unambiguous findings so far about the occurrence of giant impacts in an exoplanet system. Here, we characterize the two innermost planets of the compact and near-resonant system Kepler-107 (ref. [9]). We show that they have nearly identical radii (about 1.5-1.6 $R_\oplus$ ), but the outer planet Kepler-107c is more than twice as dense (about 12.6 g cm$^{-3}$) as the innermost Kepler-107b (about 5.3 g cm$^{-3}$). In consequence, Kepler-107c must have a larger iron core fraction than Kepler-107b. This imbalance cannot be explained by the stellar XUV irradiation, which would conversely make the more-irradiated and less-massive planet Kepler-107b denser than Kepler-107c. Instead, the dissimilar densities are consistent with a giant impact event on Kepler-107c that would have stripped off part of its silicate mantle. This hypothesis is supported by theoretical predictions from collisional mantle stripping[10], which match the mass and radius of Kepler-107c.**

The Kepler space telescope discovered the four sub-Neptune-sized planets Kepler-107b, c, d and e (ref. [9]), which transit (that is, pass in front of) their host star and have orbital periods ($P$) of 3.180, 4.901, 7.958 and 14.749 d, respectively. The four planets form a compact near-resonant system with period ratios close or almost equal to ratios of small integers, specifically $P_c/P_b$=1.541 (~3:2), $P_e/P_c$=3.009 (~3:1), $P_d/P_b$=2.503 (~5:2). The near resonances imply that the Kepler-107 planets likely formed further out and then migrated inward getting trapped into resonances during the migration process[11]. The orbital eccentricities of all four planets were determined to be low, consistent with zero[12], as expected from both dynamical stability criteria and orbit circularization times for Kepler-107b and c (see Methods). Despite the proximity to resonances, no significant transit timing variations from gravitational planet-planet interactions could be detected for any planet[13], preventing the direct determination of planetary masses from the Kepler light curve alone.

To determine the planet masses, and hence their bulk densities, we observed Kepler-107 from 2014 June 21 to 2017 April 22 with the fibre-fed high-resolution (R=115,000) HARPS-N spectrograph[14] at the Telescopio Nazionale Galileo in La Palma. Our time series with 114 radial velocities has a scatter of 6 m s$^{-1}$ and shows two significant periodicities at 4.9 d and at about 14 d (see Methods and Supplementary Figs. 5 and 6). The former is straightforwardly attributed to the Doppler signal induced on the star by the second inner planet Kepler-107c; the latter is likely related to stellar magnetic activity variations though at low level, as shown by the low amplitude of Kepler flux variations (see Methods and Supplementary Fig. 2).

From the 114 HARPS-N spectra we determined the stellar atmospheric parameters (see Table 1 and Methods). Even though the parent star is richer in iron than most of the stars that are known to host low-mass planets[15], it does not present an overabundance of iron relative to Si and Mg having solar-like Fe/Si and Mg/Si photospheric ratios (see Methods and Supplementary Table 3). To determine the stellar fundamental parameters (radius, mass and age) we performed a detailed asteroseismic analysis with all the available Kepler short-cadence data. We extracted the frequencies for individual modes of stellar oscillations from the power density spectrum of the Kepler light curve (see Methods and Supplementary Fig. 4). We modelled them with the BAyesian STellar Algorithm[16] and found $M_*=1.238\pm0.029\,M_\odot$, $R_*=1.447\pm0.014\,R_\odot$, $\rho_*=0.574\pm0.029$ g cm$^{-3}$, log$g$=4.210±0.013 (cgs), in agreement with the spectroscopic value within 1σ, and an age $t=4.29^{+0.70}_{-0.56}$ Gyr (see Methods). These parameters are consistent with those that were previously determined[17] but are more precise and accurate.

By using the constraint on the stellar density as provided by our asteroseismic analysis, we modelled the transits of the four Kepler-107 planets (Supplementary Fig. 1) and derived their radii and associated uncertainties by employing a differential evolution Markov chain Monte Carlo Bayesian technique[18] (see Table 1 and Methods). We confirm the absence of significant transit timing variations (see Methods and Supplementary Fig. 8).

In the same Bayesian framework, we modelled the HARPS-N radial velocities with four Keplerians by using the updated ephemerides from the transit fitting, and an additional sinusoid to account for the activity signal at ~14 d (see Fig. 1 and Methods). More sophisticated models to take the activity-induced correlated noise into account[19] provided very consistent results (see Methods and Supplementary Table 2). The radial velocity semi-amplitudes and derived planetary masses are listed in Table 1. Only an upper limit could be established for the mass of Kepler-107d because its Doppler signal is undetected due to its expected low mass and its orbital period longer than planets b and c.

The positions of the Kepler-107 planets in the radius-mass diagram of small planets with a bulk density measured better than $3\sigma$ are shown in Fig. 2 along with iso-composition[20] curves. The two innermost planets Kepler-107b and c have nearly equal radii, that is, $R_{p,b}$=1.536±0.025 $R_\oplus$ and $R_{p,c}$=1.597±0.026 $R_\oplus$, but dissimilar masses: $M_{p,b}$=3.51±1.52 $M_\oplus$ and $M_{p,c}$=9.39±1.77 $M_\oplus$. This means that they have different bulk densities with $\rho_{p,b}$=5.3±2.3 g cm$^{-3}$ and $\rho_{p,c}$=12.6±2.4 g cm$^{-3}$ (Table 1). According to theoretical models of planetary interiors[20], both Kepler-107b and c are consistent with a rocky composition. However, Kepler-107c appears Mercury-like in composition and has an iron fraction that is at least a factor of two greater than that of Kepler-107b. Kepler-107c's iron core makes up ~70% of its mass, with the silicate mantle making up ~30% (see Methods). Kepler-107d could also be rocky, while Kepler-107e is a mini-Neptune containing a significant fraction of volatiles in the form of a thin H/He envelope and/or water ice (see Fig. 2 and Methods).

Unlike the pair of Kepler-36 neighbouring planets[21], the difference in planetary density between Kepler-107b and c cannot be explained by atmospheric escape (or photoevaporation) caused by the high XUV stellar irradiation. This irradiation may partially or totally strip the gaseous H/He envelopes of mini-Neptunes at close distances from the star and thus may yield a diversity in exoplanet bulk densities[7]. However, it cannot be responsible for the observed dissimilar densities of Kepler-107b and c because it would have led to a higher density for Kepler-107b than for Kepler-107c (see Methods), which is at odds with our results.

Alternatively, the difference in density of the two inner planets can be explained by a giant impact on Kepler-107c that removed part of its mantle, significantly reducing its fraction of silicates with respect to an Earth-like composition[22]. The radius and mass of Kepler-107c, indeed, lie on the empirically derived collisional mantle stripping curve for differentiated rocky/iron planets[10,23] (grey dashed line in Fig. 2). Smoothed particle hydrodynamics simulations show that a head-on high-speed giant impact between two ~ 10 $M_\oplus$ exoplanets in the disruption regime would result in a planet like Kepler-107c with approximately the same mass and interior composition (see Fig. 3 and Methods). Such an impact may destabilise the current resonant configuration of Kepler-107 and thus it likely occurred before the system reached resonance. Multiple less-energetic collisions may also lead to a similar outcome[24].

Other planets with a Mercury-like composition such as K2-106b (ref. [25]) and K2-229b (ref. [3]) have been recently discovered. However, they are all the innermost planets in their systems and thus alternative mechanisms to the giant impact scenario may have given rise to their high iron content, for instance mantle evaporation for the hottest planets[3] or photophoresis and disc aerodynamic fractionation yielding a depletion of silicates in the inner regions of the protoplanetary disc[26]. These other processes cannot explain the dissimilar densities of Kepler-107b and c (see Methods). An alternative explanation might be that planet c formed closer to the parent star than planet b and afterwards crossed its orbit; if so, the iron-rich composition of Kepler-107c might actually be related to formation in the silicate-poor inner region of the protoplanetary disc[26] with no need to invoke a giant collision. However, this orbit crossing should have occurred before the dispersal of the disc, so that the resulting eccentricities could have been damped. Given the relatively large mass of planet c, it seems unlikely that there would have been sufficient time for such a scenario to operate, or that it may have occurred without the eccentricities of the lower-mass planets becoming large enough to destabilize the system.

Giant impacts are thought to have occurred in our Solar System and have been invoked to explain the composition of Mercury[27], the origin of the Earth-Moon system[28] and the high orbital obliquity of Uranus[29]. We have shown that they likely occurred in the exoplanetary system Kepler-107 and shaped the compositional properties of its two inner planets. If catastrophic disruption impacts occur frequently, we predict a clustering of small exoplanets along the maximum-collisional stripping curve[10] in the mass-radius diagram, as an increasing number of exoplanets are characterized with precise radius and mass determinations.

# Table 1: Properties of the Kepler-107 planetary system

| Parameter | Value and 68.3% credible interval |
|---|---|
| **Host star** | **Kepler-107, KIC-10875245, KOI-117, 2MASS 19480677+4812309** |
| Magnitudes[a] | $B$=13.34, $V$=12.70, $J$=11.39, $K$=11.06 |
| Distance[b] (pc) | 525.5 ± 5.5 |
| Systemic radial velocity $V_r$ (km/s) | 5.64423 ± 4.5x10$^{-4}$ |
| Effective Temperature $T_{eff}$ (K) | 5854 ± 61 |
| Metallicity [Fe/H] (dex) | 0.321 ± 0.065 |
| Spectroscopic surface gravity log$g$ (cgs) | 4.28 ± 0.10 |
| Asteroseismic surface gravity log$g$ (cgs) | 4.210 ± 0.013 |
| Rotational velocity $V\sin i$ (km/s) | 3.6 ± 0.5 |
| Mass $M_\star$ ($M_\odot$) | 1.238 ± 0.029 |
| Radius $R_\star$ ($R_\odot$) | 1.447 ± 0.014 |
| Density $\rho_\star$ (g cm$^{-3}$) | 0.574 ± 0.029 |
| Age $t$ (Gyr) | $4.29^{+0.70}_{-0.56}$ |
| Limb-darkening coefficients | $u_1$=0.26 ± 0.08, $u_2$=0.56 ± 0.12 |

| Planets | b | c | d | e |
|---|---|---|---|---|
| Orbital period $P$ (d) | 3.1800218 ± 2.9x10$^{-6}$ | 4.901452 ± 1.0x10$^{-5}$ | 7.95839 ± 1.2x10$^{-4}$ | 14.749143 ± 1.9x10$^{-5}$ |
| Transit epoch $T_c$ - 2,450,000 (BJD$_{TDB}$) | 5701.08414 ± 3.7x10$^{-4}$ | 5697.01829 ± 7.9x10$^{-4}$ | 5702.9547 ± 0.0060 | 5694.48550 ± 4.6x10$^{-4}$ |
| Transit duration (h) | 3.633 ± 0.021 | 4.269 ± 0.037 | 4.18 ± 0.36 | 6.117 ± 0.031 |
| Radius ratio $R_p/R_\star$ | 0.00973 ± 1.3x10$^{-4}$ | 0.01012 ± 1.3x10$^{-4}$ | 0.00544 ± 3.6x10$^{-4}$ | 0.01839 ± 1.4x10$^{-4}$ |
| Inclination $i$ (deg) | 89.05 ± 0.67 | $89.49^{+0.34}_{-0.44}$ | $87.55^{+0.64}_{-0.48}$ | 89.67 ± 0.22 |
| Impact parameter $b$ | 0.11 ± 0.08 | $0.08^{+0.07}_{-0.05}$ | $0.53^{+0.10}_{-0.14}$ | 0.11 ± 0.07 |
| Eccentricity $e$ | 0 (fixed) | 0 (fixed) | 0 (fixed) | 0 (fixed) |
| Radial-velocity semi-amplitude $K$ (m s$^{-1}$) | 1.32 ± 0.57 | 3.06 ± 0.57 | < 1.06 | 1.95 ± 0.82 |
| Radius $R_p$ ($R_\oplus$) | 1.536 ± 0.025 | 1.597 ± 0.026 | 0.86 ± 0.06 | 2.903 ± 0.035 |
| Mass $M_p$ ($M_\oplus$) | 3.51 ± 1.52 | 9.39 ± 1.77 | < 3.8 | 8.6 ± 3.6 |
| Density $\rho_p$ (g cm$^{-3}$) | 5.3 ± 2.3 | 12.65 ± 2.45 | < 33.1 | 2.00 ± 0.82 |
| Semi-major axis $a$ (AU) | 0.04544 ± 3.5x10$^{-4}$ | 0.06064 ± 4.7x10$^{-4}$ | 0.08377 ± 6.5x10$^{-4}$ | 0.12639 ± 9.9x10$^{-4}$ |
| Equilibrium temperature[c] (K) | 1593 ± 19 | 1379 ± 17 | 1173 ± 14 | 955 ± 12 |

[a]: B and V Johnson magnitudes from the Howell Everett Survey; J and K magnitudes from 2MASS.
[b]: From the second release Gaia data[30].
[c]: Black-body equilibrium temperature assuming a null Bond albedo and uniform heat redistribution to the night side.

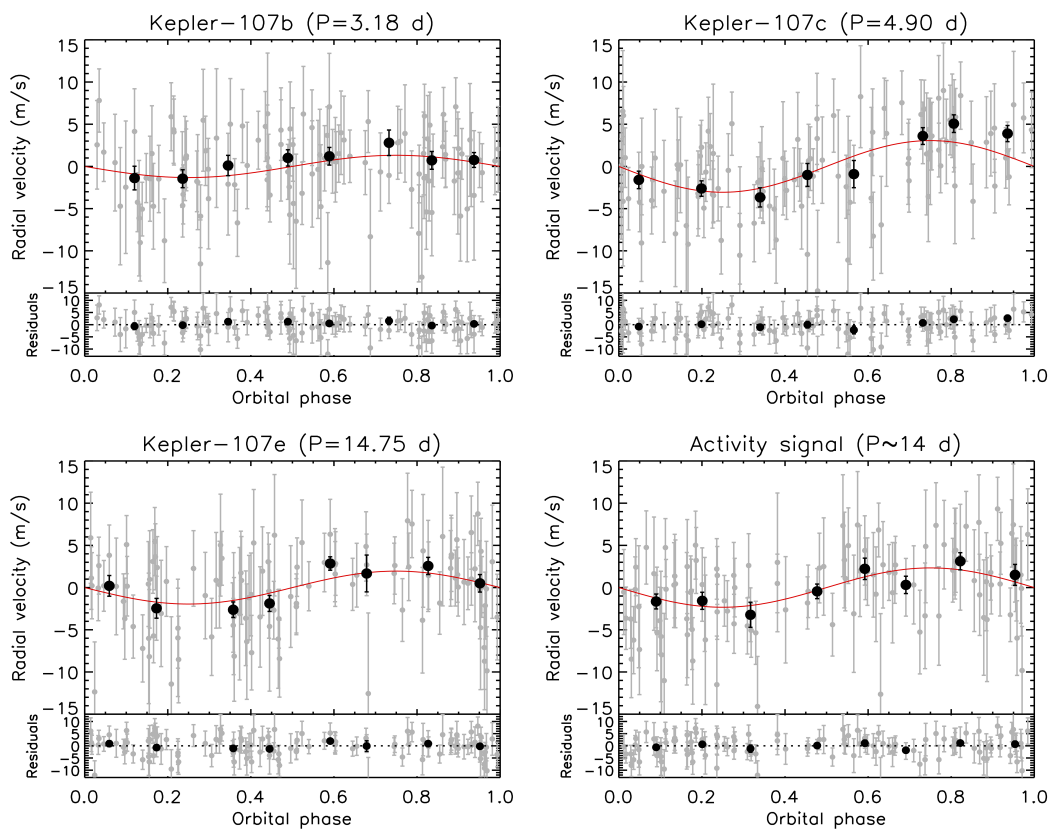

**Fig. 1 | Radial velocity measurements of Kepler-107 with HARPS-N.** Unbinned (grey dots) and binned (black circles) radial velocities with associated 1σ error bars are shown as a function of orbital phase for the planets Kepler-107b, c and e (top left, top right, and bottom left panels). The residuals after subtracting the Keplerian models (red solid lines) are also displayed. The radial velocities of Kepler-107d are not shown because they exhibit no variations. The additional signal with $P\sim14$ d, which we attribute to stellar activity, is shown on the bottom right panel.

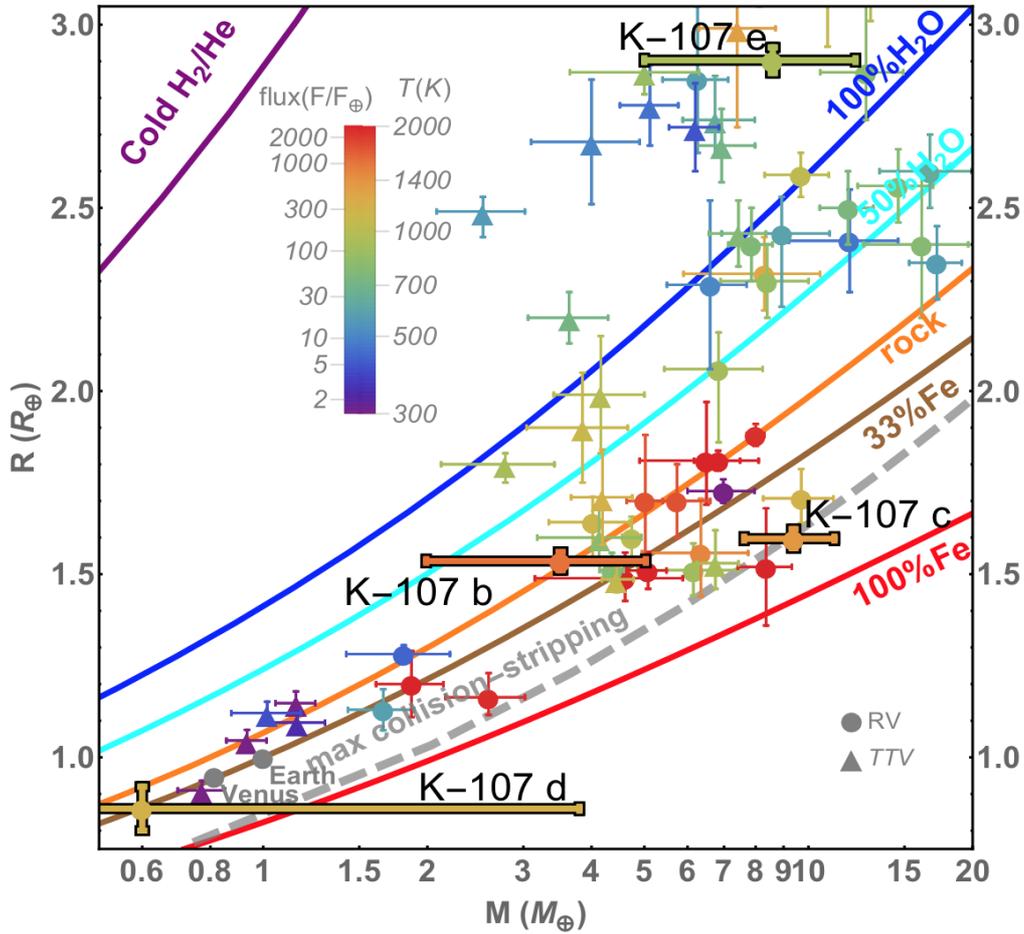

**Fig. 2 | Mass–radius diagram of exoplanets smaller than** $3\,R_\oplus$. Only known planets with bulk densities measured to better than 33% precision through radial velocities (RV, circles) and transit timing variations (TTV, triangles) are shown. Error bars represent 1σ confidence intervals. Colours of symbols are a function of the planet irradiation level (or equilibrium temperature) assuming an Earth-like albedo. The solid lines are theoretical mass-radius curves[20] for planets with different compositions as specified in the figure (see also Methods). The grey dashed line displays the minimum radius predicted from collisional stripping models as a function of planetary mass[10].

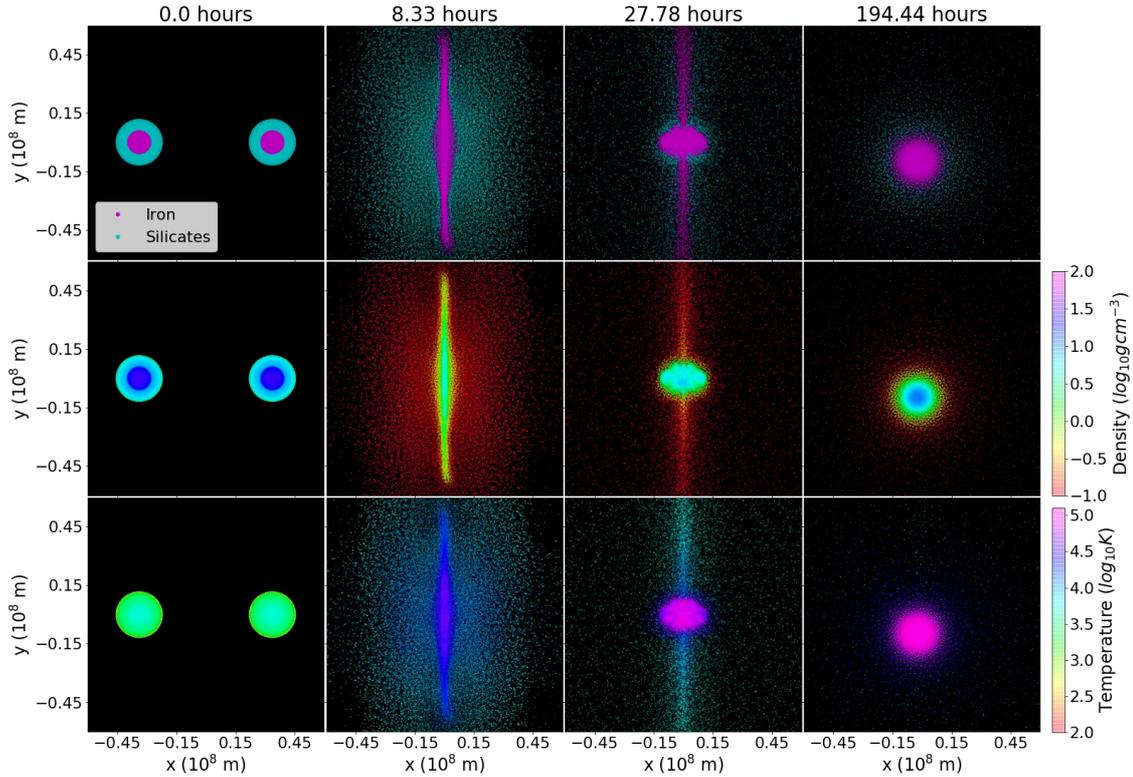

**Fig. 3 | Smoothed particle hydrodynamical collision simulation.** Hemispheric cross-sectional images of a high speed (62.5 km/s) head-on equal-mass ($10.5\,M_\oplus$, $1.8\,R_\oplus$) simulation (see Methods) are shown at four snapshots in time perpendicular to the impact plane. The strips display composition (top), density (middle) and temperature (bottom). The impactors have identical initial composition consisting of 70% forsterite mantle and 30% iron core by mass. The middle two frames show a cross-sectional cut through the pancake structure created by the impact. The majority of the material is vaporised during the impact. The largest post-collision remnant has a bound mass of $8.2\,M_\oplus$ with a material composition of 36% forsterite and 64% iron, which is consistent with both Kepler-107c mass and composition within 1σ. The post-collision remnant is significantly inflated and still cooling at the end of the simulation (last frame).

**Acknowledgments:** The authors wish to thank Dr. R. D. Haywood, Dr. R. Silvotti and Prof. D. Charbonneau for useful discussions. The HARPS-N project was funded by the Prodex Program of the Swiss Space Office (SSO), the Harvard-University Origin of Life Initiative (HUOLI), the Scottish Universities Physics Alliance (SUPA), the University of Geneva, the SmithsonianAstrophysical Observatory (SAO), the Italian National Astrophysical Institute (INAF), University of St. Andrews, Queen's University Belfast and University of Edinburgh. The present work is based on observations made with the Italian Telescopio Nazionale Galileo (TNG) operated on the island of La Palma by the Fundación Galileo Galilei of the INAF (Istituto Nazionale di AstroFisica) at the Spanish Observatorio del Roque de los Muchachos of the Instituto de Astrofísica de Canarias. This paper exploited data collected by the Kepler mission; funding for the Kepler mission is provided by the NASA (National Aeronautics and Space Administration) Science Mission directorate. The research leading to these results received funding from the European Union Seventh Framework Programme (FP7/2007-2013) under grant agreement number 313014 (ETAEARTH). L.Z. acknowledges support from the Simons Foundation (SCOL[award #337090]). MD acknowledges financial support from Progetto Premiale 2015 FRONTIERA funding scheme of the Italian Ministry of Education, University, and Research. Funding for the Stellar Astrophysics Centre is provided by The Danish National Research Foundation (Grant agreement No. DNRF106). V.S.A. acknowledges support from VILLUM FONDEN (research grant 10118). E.C. is funded by the European Union's Horizon 2020 research and innovation program under the Marie Sklodowska-Curie grant agreement No. 664931. T.D. is supported by a STFC PhD studentship. R.A.G. acknowledges support from CNES. This work has been carried out in the frame of the National Centre for Competence in Research PlanetS supported by the Swiss National Science Foundation (SNSF). C.L., F.B., F.P. and S.U. acknowledge the financial support of the SNSF. M.S.L. is supported by The Independent Research Fund Denmark's Sapere Aude program (Grant agreement No. DFF–5051-00130). S.M. acknowledges support from the Ramon y Cajal fellowship number RYC-2015-17697.

**Author contributions:** The underlying radial-velocity observation programme was conceived and organized by F.P., A.C.C., D.W.L., C.L., D. Ségransan, S.U. and E.M.. Observations with HARPS-N were carried out by L.A., A.C.C., A.M., C.D.D., M.D., X.D., R.C., A.F.M.F., P.F., A.H., F.M., M.L.M., L.M., C.N., V.N., K.R. and A.V.. C.L. maintained and updated the reduction pipeline, L.M. implemented the correction of radial velocities for moonlight contamination, and X.D. computed the values of the $\log(R'_{HK})$ activity indicator. A.S.B., M.D. and K.R. analysed and modelled the radial velocities. L.M. simulated and compared the radial velocities using both non-interacting and interacting Keplerians. A.S.B. also performed the transit fitting and S.M.M. worked on the analysis of transit timing variations. A.S.B. and A.V. analysed the Kepler light curve in search of the stellar rotational modulation signal. L.B. and A.M. determined the stellar atmospheric parameters from the HARPS-N spectra; A.M. also derived the stellar chemical abundances. T.A., M.B., E.C., J.C.D., R.A.G., R.H., A.B.J., H.K., M.N.L., M.S.L., S.M., V.S.A. carried out the asteroseismic analysis of the Kepler light curve and determined the stellar parameters from which A.S.B. and M.D. derived the planetary orbital and physical parameters. L.Z. deduced the planet interior compositions and E.L. estimated the planet atmospheric escapes. K.R. carried out dynamical simulations of the planetary system, and Z.M.L. and T.D. conducted the simulations of giant impact events. A.S.B. was the primary author of the manuscript for which received important contributions by L.Z., Z.M.L., M.D., E.L., M.N.L., V.S.A., A.S., A.M. and M.L.M.. All authors have contributed to the interpretation of the data and the results.

**Correspondence and requests for materials** should be addressed to A.S.B.

# METHODS

**Kepler light curve, transit detection and stellar variability.**
We downloaded the Kepler light curves extracted by both the Simple Aperture Photometry (SAP) and Pre-search Data Conditioning (PDC) pipelines with long-cadence (29.42 min) and short-cadence (58.8 s) sampling[31]. Photometric data were gathered over 4 years, from 2009 May 2 to 2013 May 11, in Kepler quarters (i.e., three-monthly intervals) 0-6, 8-10, 12-14, 16-17 in long cadence and, starting from quarter 3, in short cadence as well. We performed a low-pass filter of the PDC short-cadence light curve and searched for transits following ref. 32; we could thus confirm the previously detected transit signals[9] and found no additional transiting companions.

The long-cadence PDC light curve with transits removed shows a peak-to-peak amplitude of ~0.1%, with a maximum of ~0.2% between 1100 and 1300 $BJD_{TDB}$-2,454,900 (Supplementary Fig. 2). We performed a weighted autocorrelation function of this light curve finding the highest, though low-amplitude, peak at ~14 d (Supplementary Fig. 3). This peak might be the stellar rotation period $P_{rot}$, which would be consistent with the upper limit given by the projected rotational velocity $V\sin i$ (Table 1) assuming an edge-on stellar equator, that is, $P_{rot} < 20.3^{+3.3}_{-2.5}$ d ($P_{rot} < 20.3^{+14.5}_{-6.0}$ d) at 1σ (3σ). No additional information on $P_{rot}$ could be extracted from the stellar activity indicators (see below).

**Asteroseismic analysis of Kepler data.**
To perform the asteroseismic analysis on all the available short-cadence Kepler data, we first corrected them using the KASOC[33] and KADACS[34] filter pipelines. We then computed the power density spectrum from a weighted least-squares sine wave fitting, which was single-sided calibrated, normalized according to Parseval's theorem, and converted to power density by dividing by the integral of the spectral window[35].

We performed a Bayesian peak-bagging analysis to extract the frequencies for individual modes of oscillation, following the methodologies outlined in refs. 36-38. To define the priors for the mode frequencies, we derived initial guesses by hand from smoothed versions of the power density spectrum using the estimated value for the large frequency spacing ($\Delta\nu$), together with an approximate value for the dimensionless offset ($\varepsilon$) for a proper mode identification[39]. Given the low signal-to-noise ratio (S/N) of the oscillation power, we imposed Gaussian priors on the oscillation mode linewidths based on the empirical relation for linewidth as a function of frequency given in ref. 39; if uninformative priors are used at the S/N observed for this star, one simply risks fitting noise spikes near the initial guess for the mode frequency. Following the procedure described in ref. 40, we calculated for each fitted mode a metric for the quality of the fit which was adopted in the stellar modelling efforts to decide on the number of modes to include in the analysis. Supplementary Fig. 4 shows the power density spectrum of Kepler-107 together with the best fitting model spectrum from the peak bagging.

**Stellar parameters.**
Our HARPS-N spectra were reduced with the online Data Reduction Software and used to determine the stellar atmospheric parameters, that is $T_{eff}$, log$g$, [Fe/H], with two different techniques: the first one relies on the measure of the equivalent widths of a set of iron lines[41], while the second one compares the observed spectra with a library grid of synthetic template spectra[15]. The two methods provided fully consistent parameters and we took the averages of their values and error bars as the final parameters and uncertainties (Table 1). The latter technique also allowed us to estimate the stellar projected rotational velocity $V\sin i$=3.6±0.5 km s$^{-1}$. Moreover, we derived the stellar abundances relative to the Sun (Supplementary Table 3) by using the same method as in ref. 42 and the solar abundances in ref. 43.

We determined the host star physical parameters, that is density, radius, mass and age (Table 1), by fitting the peak-bagged oscillation frequencies to asteroseismic predictions from stellar evolution models with the BAyesian STellar Algorithm[16]. Briefly, we constructed a fine grid of stellar tracks using the GARching STellar Evolution Code[44] and computed the associated theoretical frequencies of oscillation with the Aarhus adiabatic oscillation package[45]. The adopted input physics includes microscopic diffusion of helium and heavy elements following prescription and overshooting given in ref. 46 and based on the model by ref. 47. To avoid the influence of the so-called surface effect in our calculations[48], we chose to reproduce ratios of oscillation frequencies that have been shown to minimize the influence of the outermost stellar layers (see, for example, ref. 49 and references therein). The derived stellar radius, $R_\star = 1.447 \pm 0.014\,R_\odot$ agrees within $1\sigma$ with that determined from the Gaia parallax[30], that is $R_\star = 1.45 \pm 0.06\,R_\odot$, but is significantly more precise.

**Radial-velocity observations and modelling.**
In total we gathered 120 HARPS-N spectra by using typical exposure times of 30 min. The radial velocities (RVs) were extracted from the HARPS-N spectra by performing a weighted cross correlation with a numerical spectral mask of a G2V star[50] and have a median uncertainty of 4.5 m s$^{-1}$. Twenty-two RVs were contaminated by moonlight and were corrected with the method described in ref. 51. Four measurements were discarded because of low S/N, which yields RV uncertainties larger than 10 m/s; two more RV points were identified as outliers on the basis of Chauvenet's criterion and also excluded. Our RV time series thus contains 114 RV measurements that passed these criteria, and is shown in Supplementary Fig. 5. The RVs, their epochs and uncertainties are listed in Supplementary Table 4 along with the measurements of activity indicators such as the full-width at half maximum (FWHM) and the bisector (BIS) of the averaged line profile[52], which were computed from the cross-correlation function, and the logarithm of the $R'_{HK}$ parameter defined as the ratio of the chromospheric emission in the cores of the CaII H and K lines to the stellar bolometric emission[53].

We searched for periodic signals in the RV and activity indicator time series by using Generalised Lomb-Scargle (GLS) periodograms[54] and taking the measurement uncertainties into account. The obtained power spectra along with the false alarm probabilities[54] and the spectral window are shown in Supplementary Fig. 6. In the RV time series the most significant periodicities are found at $4.899 \pm 0.003$ d and $13.79 \pm 0.03$ d. The peak at 4.9 d corresponds to the orbital period of Kepler-107c as determined from the observation of its transits in the Kepler light curve, and this RV signal is indeed in phase with the orbital ephemeris of Kepler-107c (see top-right panel in Fig. 1) as we expect if it is planetary in origin. The peak at 13.8 d is accompanied by a close peak at $14.30 \pm 0.03$ d which falls at the secondary peak of the spectral window; therefore, one of the two close peaks is the alias of the other. Regardless of the ambiguity, this periodicity at $\sim 14$ d is likely related to stellar activity because it does not correspond to any of the transit periods and is compatible with the highest peak in the autocorrelation function of the Kepler light curve (Supplementary Fig. 3). Other peaks that are just above the significance threshold in Supplementary Fig. 6 (top panel) disappear as soon as the 4.9 d and 13.8 d signals are removed from the original dataset and are thus aliases as well. No significant periodicities are found in the activity indicator measurements at the RV peaks (Supplementary Fig. 6) and there are no significant correlations between activity indices and RVs. In general, the host star is not active with a median value of $\log(R'_{HK})$ equal to $-5.08 \pm 0.14$ dex.

To determine the RV semi-amplitudes of the four Kepler planets and hence their masses, we modelled the RVs with four non-interacting Keplerians, an additional sinusoid to take the activity signal at $\sim 14$ d into account, the systemic radial velocity, and an RV uncorrelated jitter term to account for extra noise as in ref. 55. We first adopted circular orbits for all the four planets; indeed, the orbit circularization times of the innermost planets Kepler-107b and Kepler-107c due to tidal dissipation inside the planets[56] are shorter than the system age, $t = 4.29^{+0.70}_{-0.56}$ Gyr. Specifically, by assuming an Earth-like modified tidal quality factor $Q'_p \approx 1500$ (ref. 57) which is reasonable for rocky planets (and likely overestimated for Mercury-like planets), we find circularization times of $\sim 0.13$ and $\sim 1.7$ Gyr for planets b and c, respectively. We also adopted circular orbits for Kepler-107d and Kepler-107e in the absence of any useful constraint on the orbital eccentricity from RVs, given the non-detection of the Kepler-107d Doppler signal and the low-amplitude of that of Kepler-107e. This is a reasonable assumption given that N-body dynamical simulations we ran for 10 Myr, using the Mercury6 code[58] with the masses shown in Table 1, indicate that the eccentricities of planets d and e must be lower than 0.15 to achieve dynamical stability. Eccentricities consistent with zero for all four planets were also found from Kepler photometry uniquely[12].

The posterior distributions of the 17 free parameters were obtained in a Bayesian framework by employing a differential evolution Markov chain Monte Carlo tool (DE-MCMC), following the prescriptions given in refs. 18 and 59 and, in particular, running 34 chains (twice the number of free parameters). Uninformative priors with reasonably large bounds were imposed on all the parameters except for the transit mid-times and periods of the four planets for which Gaussian priors were set from Kepler photometry (Supplementary Table 1). The medians and the 15.86% and 84.14% quantiles of the posterior distributions were taken as the best values and 1σ uncertainties, and are reported in Table 1 and Supplementary Table 2. Very consistent results on orbital parameters and planetary masses are achieved when including the six discarded RVs and/or without performing the moonlight correction.

We also fitted the RVs by employing a Gaussian process (GP) regression[60] with a quasi-periodic kernel to model the contribution due to the stellar variability following ref. 19. We used the MultiNest v3.10 Bayesian inference tool[61] with the same implementation as in ref. 62 to sample the parameter space and derive the posterior distributions for the 18 model parameters. The latter are the orbital parameters (period, transit time, RV semi-amplitude) of the four Kepler-107 planets, the systemic RV, the uncorrelated jitter term and the four GP covariance hyper-parameters, that is, the amplitude $h$ of the correlations, the periodicity $\theta$ which can be related to the stellar rotation period, the length scale $w$ of the periodic component and the correlation decay time scale $\lambda$, which can be associated with the lifetime of active regions (see equation 1 in ref. 19). The adopted prior distributions for each fitted parameter are listed in Supplementary Table 1. We used 800 random walkers (live points) and set the sampling efficiency parameter to 0.5. The RV semi-amplitudes and their 1σ uncertainties obtained with both the GP analysis and the sinusoidal-activity model are fully consistent (Supplementary Table 2). However, some of the GP hyper-parameters, such as the $\lambda$, $w$ and, to a lesser extent, $h$ parameters are not well constrained. Nonetheless, we note that the $\theta$ parameter is quite well confined, close to the expected ~14 d activity signal periodicity, even if the used uniform prior interval was large enough, that is [12, 16] d.

Bayesian model comparison between the sinusoidal-activity model and that employing GP regression was performed using MultiNest[61] and the same priors as for the latter model (third column in Supplementary Table 1). The obtained Bayes' factor of ~3.4 in favour of the sinusoidal-activity model does not support the choice of the more complex GP model[63].

To evaluate the possible effects of low but non-zero eccentricities on the recovered RV semi-amplitudes and hence on the planetary masses, we re-did the DE-MCMC orbital fit by letting the eccentricity of the four planets vary, though with reasonable priors. Specifically, we imposed Gaussian priors with zero mean and standard deviation $\sigma_e=0.05$ on the eccentricities of Kepler-107b and c, and Rayleigh priors with $\sigma_e=0.04$ for the two outer planets[64]. The derived RV semi-amplitudes $K_b=1.32\pm0.57$, $K_c=3.08\pm0.58$, $K_d<1.08$ and $K_e=1.89\pm0.83$ m s$^{-1}$ are indistinguishable from those obtained with circular orbits (Table 1).

As a further investigation, we checked that the use of non-interacting Keplerians is accurate enough. To this end, we compared the planetary Doppler signals computed with both non-interacting and interacting Keplerians using the TRADES code[65]. The maximum difference between the two models is about 5 cm s$^{-1}$ and is thus completely negligible.

The mass of Kepler-107c, $M_{p,c}$, has been determined with a relative uncertainty of 19% (better than 5σ), while that of Kepler-107b ($M_{p,b}$) with an uncertainty of 43% given that photon-noise RV error bars are on average ~3.5 times higher than the RV semi-amplitude of planet b. According to these uncertainties, the probability that $M_{p,b}$ and $M_{p,c}$ would differ by chance is 0.6% and 0.1% for the sinusoidal-activity and GP models, respectively. To further show that $M_{p,b}$ must be considerably lower than $M_{p,c}$ despite its low precision, we simulated artificial RV time series by (1) injecting at the observation epochs the Doppler signal of Kepler-107b for three different mass values: $M_{p,b}=M_{p,c}$ ( $9.4\,M_\oplus$ ), $M_{p,b}=0.5\,M_{p,c}$ ( $4.7\,M_\oplus$ ) and $M_{p,b}=0.33\,M_{p,c}$ ( $3.1\,M_\oplus$ ); (2) adding the Doppler signals of planets c and e as well as the 13.8 d activity signal with their semi-amplitudes as reported in Table 1 and Supplementary Table 2; and (3) shifting each RV point generated from the previous steps, following a normal distribution with mean and standard deviation equal to the RV value and its 1σ error bar. In this way, for every mass value of Kepler-107b we simulated 500 RV time series with different noise realisations and computed their GLS periodogram. The averaged periodograms for each simulated $M_{p,b}$ value are displayed in Supplementary Fig. 7. These artificial periodograms clearly show that $M_{p,b}$ is lower than $M_{p,c}$ by at least a factor of ~2, as found by the Bayesian analyses, otherwise the power at $P_b$ would have been greater than observed (top panel), and even higher than the peak at $P_c$ for $M_{p,b}=M_{p,c}$ (second top panel).

**Transit fitting and planetary parameters.**
To perform the transit fitting, we treated the SAP short-cadence light curve as in ref. 66 and used the transit model of ref. 67. For each planet we fitted for the transit epoch, the orbital period, the transit duration, the radius ratio $R_p/R_\star$, the orbital inclination, and the limb-darkening coefficients $q_1=(u_1+u_2)^2$ and $q_2=0.5\,u_1/(u_1+u_2)$ (ref. 68), where $u_1$ and $u_2$ are the coefficients of the limb-darkening quadratic law[69]. We used circular orbits for all the planets as for the RV modelling. We imposed a Gaussian prior on the stellar density following, for example, ref. 70, by taking advantage of the very precise and accurate value derived from the asteroseismic analysis of Kepler data ( $\rho_\star=0.574\pm0.029$ g cm$^{-3}$ ). No bounds were set on the transit parameters except for the $q_1$ and $q_2$ parameters for which intervals of [0, 1] were adopted[68].

The posterior distributions of the free parameters were obtained with the same DE-MCMC technique as above and were used to derive the medians and 1σ confidence intervals, which are reported in Table 1 except for $q_1$ and $q_2$ which were found to be 0.67 ± 0.10 and 0.56 ± 0.12, respectively. The derived radii of planets b, c and e are fully consistent (within 1σ) with those which were previously determined in ref. 12; for planet d we found a slightly smaller radius. The planetary orbital and physical parameters were determined by combining the posterior distributions of the stellar, RV and transit parameters, and are also listed in Table 1.

Moreover, we computed the transit timing variations (TTVs) for each planet with the short-cadence data in the same Bayesian framework. Because of the low S/N of the individual transits, in particular for planets b, c and d, we fitted to each transit the model obtained with the whole light curve by letting only the mid-transit time vary. The Observed-Calculated (O-C) transit times of planets b, c and e showing no significant variations from a constant ephemeris are displayed in Supplementary Fig. 8. Those of planet d could not be determined because of the very low S/N of the individual transits (see Supplementary Fig. 1, bottom-left panel), preventing the DE-MCMC chains from achieving proper convergence in a single transit fitting, and are not shown.

Our non-detection of transit timing variations is in agreement with the results of ref. 13. Using the formalism from ref. 71, one can estimate the expected TTV amplitudes due to neighbouring planets with $e\approx0$. Planet b's TTV amplitude is expected to be ~3 min due to planet c; the TTV amplitude of planet c is ~2 min from planet b and also ~2 min from planet d, using the RV measured (or upper limit) masses. The interior planets have average TTV uncertainties of 25 min and hence the non-detection of TTVs is not surprising. We also ran a photodynamic MCMC[72] to determine the upper limits of the masses allowed with the observed transit data. We find upper limits of 22, 24, 8.6 and $22\,M_\oplus$ for planets b, c, d and e at the 1σ level, which are all significantly higher than the masses and upper limits measured via RVs (Table 1).

**Planetary composition.**
The equation-of-state data allow calculations of planet average densities under self-gravitational compression and further constrain planet bulk compositions and internal structures following the methods in ref. 20. The Earth-like composition is assumed to be 32.5 wt.% (weight percent) Fe/Ni-metal plus 67.5 wt.% MgSiO$_3$-rock, fully differentiated into a bi-layer core-mantle structure. 32.5% of Fe/Ni-metal by mass is consistent with both the general cosmic element abundance (number) ratio of Mg:Si:Fe being close to 1:1:1 and the Fe/Si and Mg/Si ratios derived from the host star abundances (Supplementary Table 3), that is, Fe/Si= $0.89^{+0.20}_{-0.17}$ and Mg/Si= $1.22^{+0.25}_{-0.21}$, which can be considered as a proxy for the protoplanetary disc composition. According to these ratios, the iron-rich composition of Kepler-107c cannot be primordial[73]. Moreover, Fe/Ni-metal and Mg-silicates have similar volatility and condensation temperatures[74] in nebula; therefore, it is rather difficult to alter these ratios in the planet composition through merely temperature effects in the protoplanetary disc, and in particular for the Kepler-107 system where the outer planet (c) is denser than the inner one (b). A special scenario such as a giant impact must be invoked to account for that.

Together, Fe/Ni-metal and Mg-silicates comprise about half a percent of the nebula mass in a solar-metallicity nebula. Differentiation is expected to occur on all rocky objects during their formation due to melting or partial melting from the energy of early radioactive elements and accretion heat. $H_2O$ is assumed to be in solid form along the melting curve (liquid-solid phase boundary). The 50% $H_2O$ curve in Fig. 2 corresponds to an Earth-like composition with exactly equal mass of $H_2O$ on top. The ices that condensed out of the nebula are expected to be a mixture of $H_2O$-$NH_3$-$CH_4$. The condensation of ices is expected to occur near icelines. If all $H_2O$ ice condenses out of the gas phase, it is about equal mass as Fe/Ni-metal plus Mg-silicates, that is, another half a percent of the uncondensed nebula by mass. And if $NH_3$-$CH_4$-clathrate-ices all condense out, it is yet another half a percent of the uncondensed nebula by mass. The mass-radius curve of $H_2O$-$NH_3$-$CH_4$ mixture is expected to be very similar to that of pure $H_2O$. The $H_2$-He gaseous content is assumed to be a cosmic mixture of 75% $H_2$ and 25% He by mass. Changing the metallicity of the host star and thus of the nebula, given that the star Kepler-107 is metal-rich, will change the total amount of condensed materials with respect to $H_2$-He gas, but it will not change the relative mass ratio of Fe metal versus Mg silicates, $H_2O$ ice, $NH_3$-$CH_4$-clathrate-ices, as this ratio is ultimately dictated by the cosmic nuclear synthesis processes occurring in giant star interiors and supernovae over the history of our Milky Way galaxy.

**Planetary evolution and escape simulations.**
Planetary evolution models were run using the coupled structure, thermal evolution and photoevaporation model of ref. 75. As the inner planets in the system are all consistent with bare rocky compositions, while the mass and radius of planet e require that it has a significant gaseous envelope, for our simulations we assumed that these planets initially formed with solar composition hydrogen and helium envelopes atop rock and iron cores. We then calculated the mass, radius and composition evolution of each planet due to thermal cooling and atmospheric escape from XUV-driven photoevaporative escape.

For a given planet composition and a given star, a planet's vulnerability to photoevaporative escape scales roughly as $F_p M_{core}^{-2}$ (ref. 21), where $F_p$ is the incident bolometric flux that a planet receives from its parent star and $M_{core}$ is the planet core mass. As a result, the relatively low masses and high irradiations of planets b and d mean that they are the most vulnerable to atmospheric escape, followed by planet c. At ~8.6 $M_\oplus$ and 14.75 d, however, planet e is massive enough and far enough out that it is not nearly as vulnerable to photoevaporation, consistent with the fact that it appears to have retained a gaseous envelope. Assuming an Earth-like core, we estimate that Kepler-107e should have ~3% of its mass in a gaseous H/He envelope today and that only had ~5% when it initially formed. On the other hand, due to its higher irradiation Kepler-107c could have easily lost a ~5% envelope in its first Gyr and planets b and d significantly more than that.

While photoevaporation may explain the overall density contrast between the inner three planets and planet e, it cannot explain the contrast between planets b and c. Planet b is more than an order of magnitude more vulnerable to photoevaporation than c, mostly because of its lower mass, and yet it is planet c that appears to be denser and much more iron-rich. Additionally, we note that at just ~1400 K, Kepler-107c is below the vaporization temperature of most silicate species[76], suggesting that vaporization and escape of the silicate mantle is unlikely to be responsible either. Indeed, silicate atmospheres do not experience any significant escape until they reach temperatures >2000

K (ref. 77), hotter than any of the planets in this system. Moreover, while the exact scaling may be different for other processes such as outgassing and escape from silicate atmospheres[77] or minor impact erosion[78,79], all escape processes should be more effective for planet b than c, as it is both more irradiated and less gravitationally bound. This qualitative fact alone suggests that the high density of planet c is unlikely to be due to a gradual escape process that affected all planets in the system consistently and must instead be due to a stochastic mechanism such as a giant impact.

**Giant impact simulations.**
A range of smoothed particle hydrodynamical simulations were run at various impact speeds using a modified version of GADGET2[80]. We focused on head-on and nearly head-on equal-mass disruptive events because these are generally the most efficient at stripping outer layers[22]. The thermodynamic properties of the different materials were calculated at each step from a tabulated equation of state[23,81], and the initial thermodynamic profiles were determined following ref. 82. Each initial impactor had a resolution of $10^5$ particles. The post-collision bound mass was calculated by taking the particle closest to the potential minimum as a seed, then iteratively calculating which additional particles were gravitationally bound to the current set. This process continued until the mass difference between iterations was below a $10^{-7}$ set tolerance[83]. We find that we can significantly change the mantle-core ratio of the target as a result of a single high-energy impact, an example of which is shown in Fig. 3. Such an impact could take a planet that originally was dominated in mass by mantle (something like planet b) and convert it into a planet dominated in mass by core (similar to planet c). Multiple energetic collisions may yield a similar outcome[24] but, given the large variety of possible configurations, their simulation goes beyond the scope of the present work.

**Data availability**
The RV data that support the findings of this study and have been used to produce some of the plots are available in the Supplementary Information. Kepler data are available at the Mikulski Archive for Space Telescopes (https://archive.stsci.edu/kepler/).

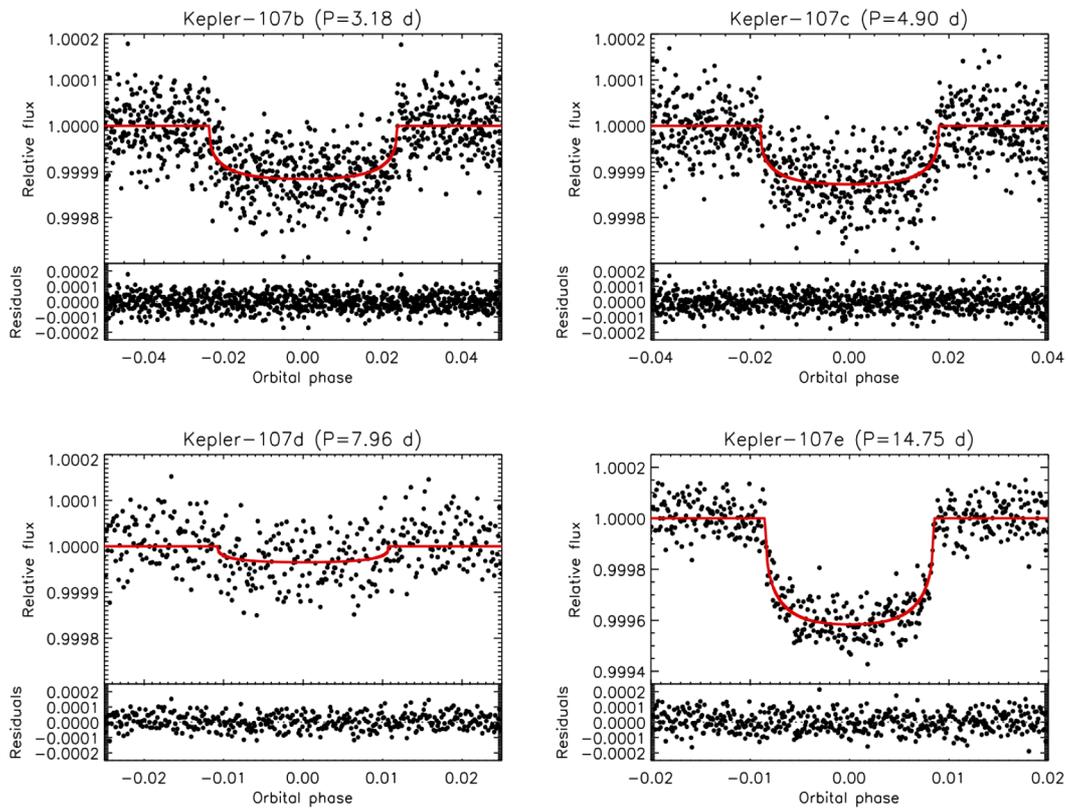

Supplementary Figure 1: **Transits of the Kepler-107 planets.** Phase-folded transits of the four Kepler-107 planets with the best-fit model (red solid line) and residuals. For illustration purposes, data were binned in phase bins of 30, 45, 85, and 110 s for planets b, c, d, and e.

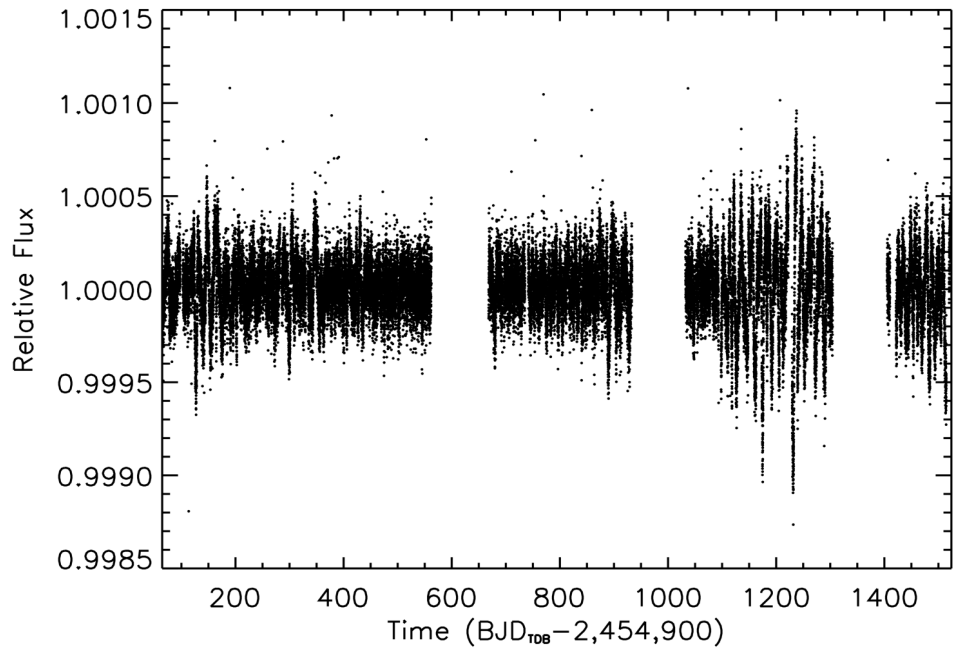

Supplementary Figure 2: **Kepler light curve.** Kepler long-cadence photometric measurements showing low-amplitude variations due to the rotational modulation of small photospheric active regions.

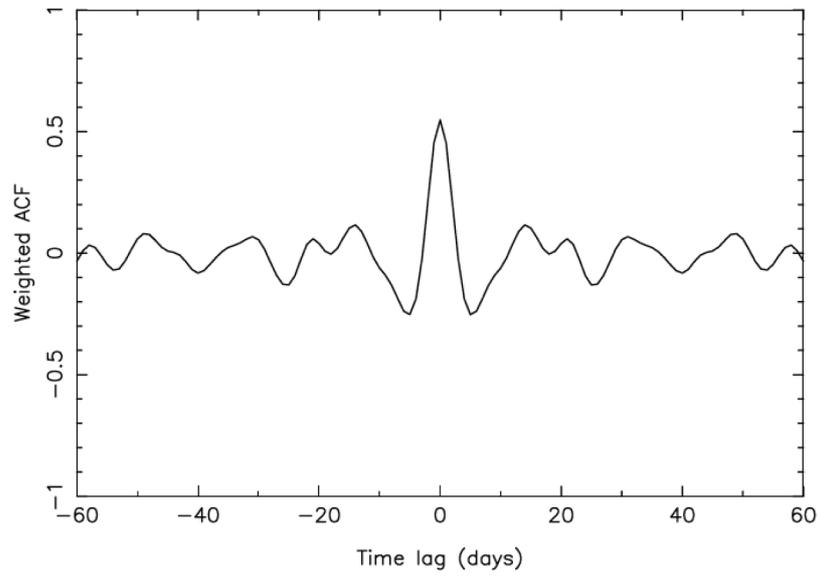

Supplementary Figure 3: **Weighted autocorrelation function of the Kepler light curve.** The peak at ~14 d might be the stellar rotation period.

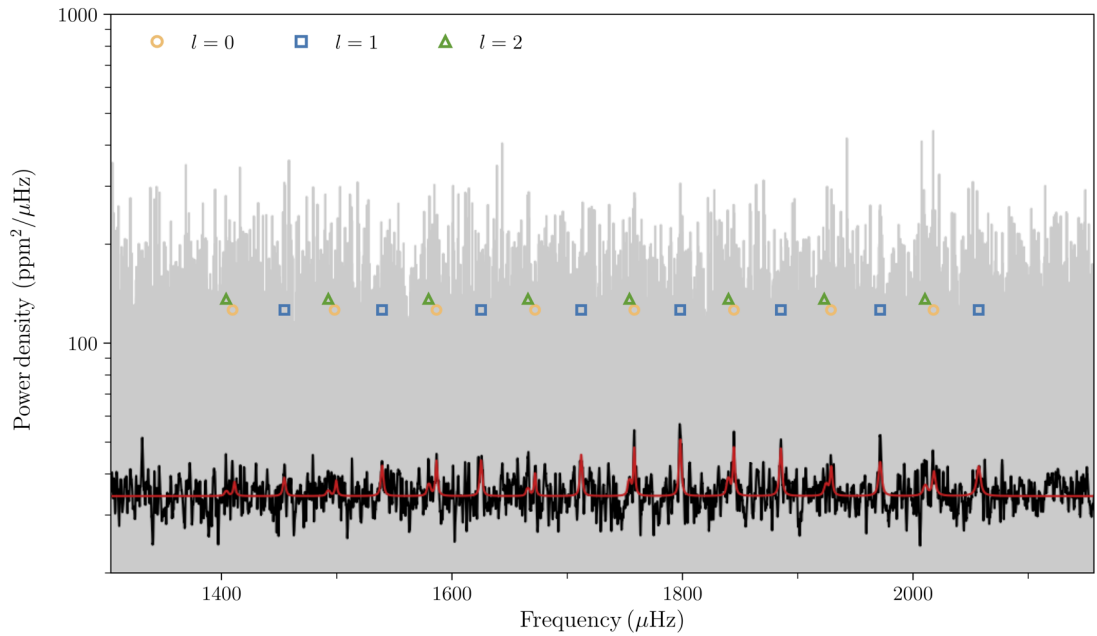

Supplementary Figure 4: **Peakbagging fit for Kepler-107.** The power density spectrum centred on the region of excess power from solar-like oscillations is shown in grey, with a 1 μHz smoothed version overlain in black. The best fitting model from the peakbagging is shown in red. The frequencies and angular degree of the individual modes are indicated by the coloured markers.

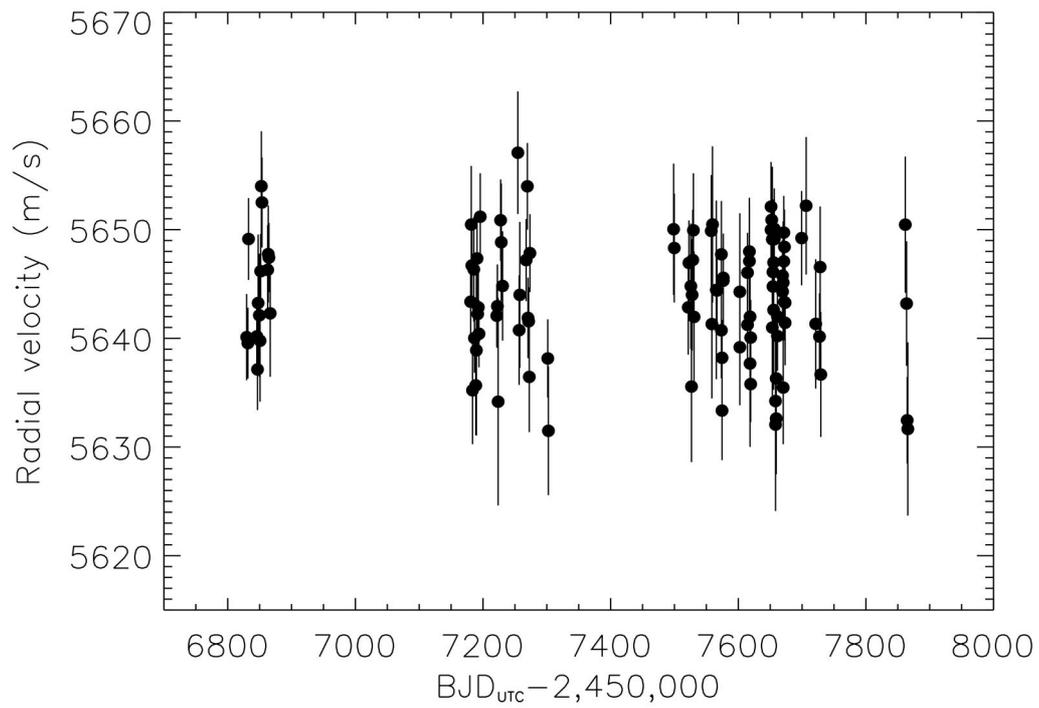

Supplementary Figure 5: **HARPS-N radial-velocity time series.** Radial velocities gathered with the HARPS-N high-accuracy and high-precision spectrograph at the Telescopio Nazionale Galileo (La Palma, Spain) as a function of time.

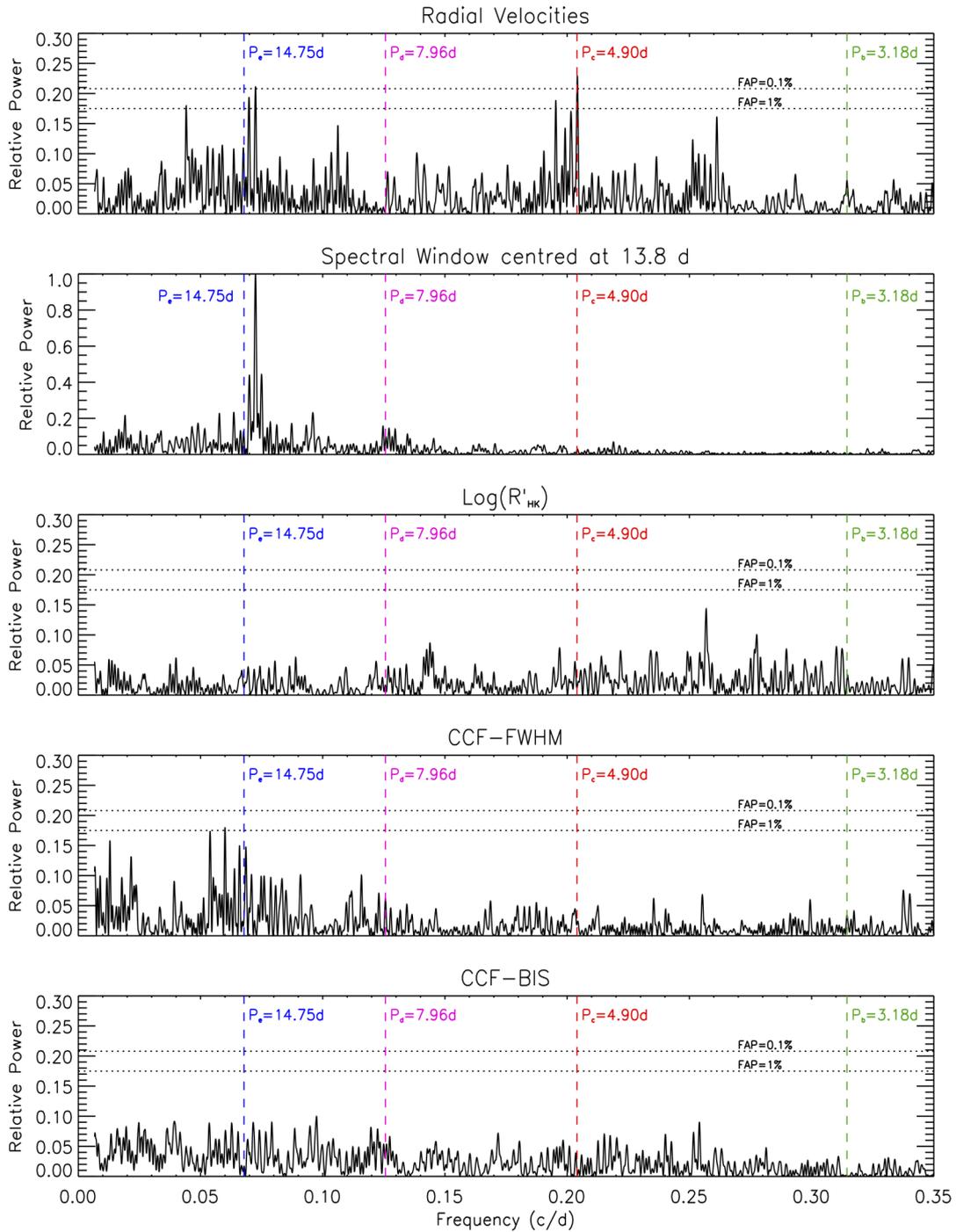

Supplementary Figure 6: **Generalized Lomb-Scargle periodograms of the radial-velocity and activity indicator time series measured by HARPS-N.** The panels show, from top to bottom, the periodograms of the RV time series, the spectral window centred at the highest peak at 13.8 d in the RV, the periodograms of the activity indexes log($R'_{HK}$), full-width at half maximum (FWHM) and bisector (BIS) of the averaged line profile. Vertical dashed lines indicate the orbital periods of Kepler-107b (green), c (red), d (pink), and e (blue). The horizontal dotted lines show the theoretical false alarm probabilities of 0.1% and 1%.

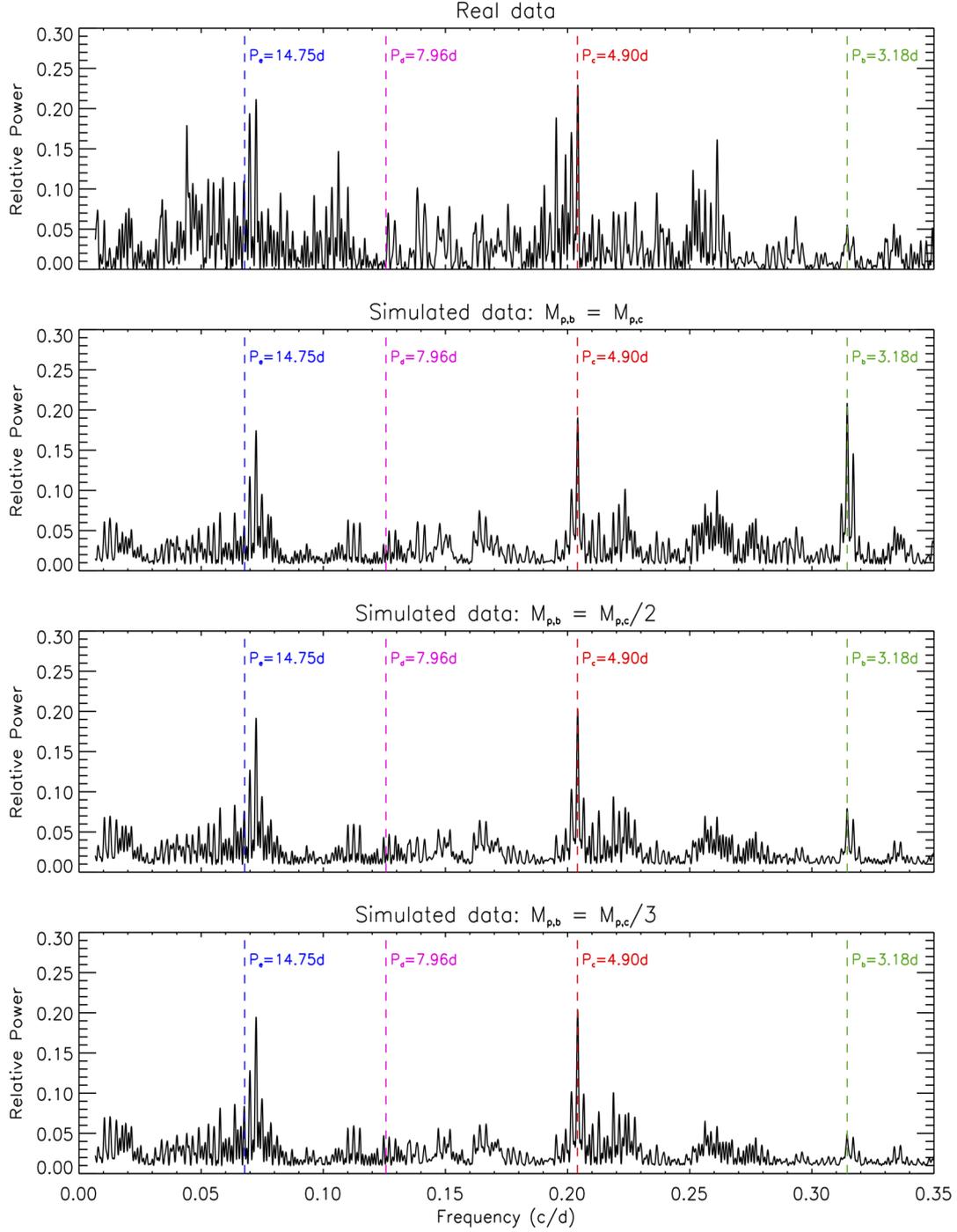

Supplementary Figure 7: **Generalized Lomb-Scargle periodograms of real and simulated radial-velocity time series.** The top panel shows the periodogram of the HARPS-N RV data (the same as in previous figure). The other panels display, from top to bottom, the averaged periodograms of simulated RVs at the real observation epochs by assuming three different values for the mass of Kepler-107b (see Methods): $M_{p,b}=M_{p,c}$ ( $9.4\,M_\oplus$ ), $M_{p,b}=0.5\,M_{p,c}$ ( $4.7\,M_\oplus$ ), and $M_{p,b}=0.33\,M_{p,c}$ ( $3.1\,M_\oplus$ ). These periodograms are less noisy than the real one (top panel) mainly because of the averaging effect. Vertical dashed lines indicate the orbital periods of Kepler-107b (green), c (red), d (pink), and e (blue).

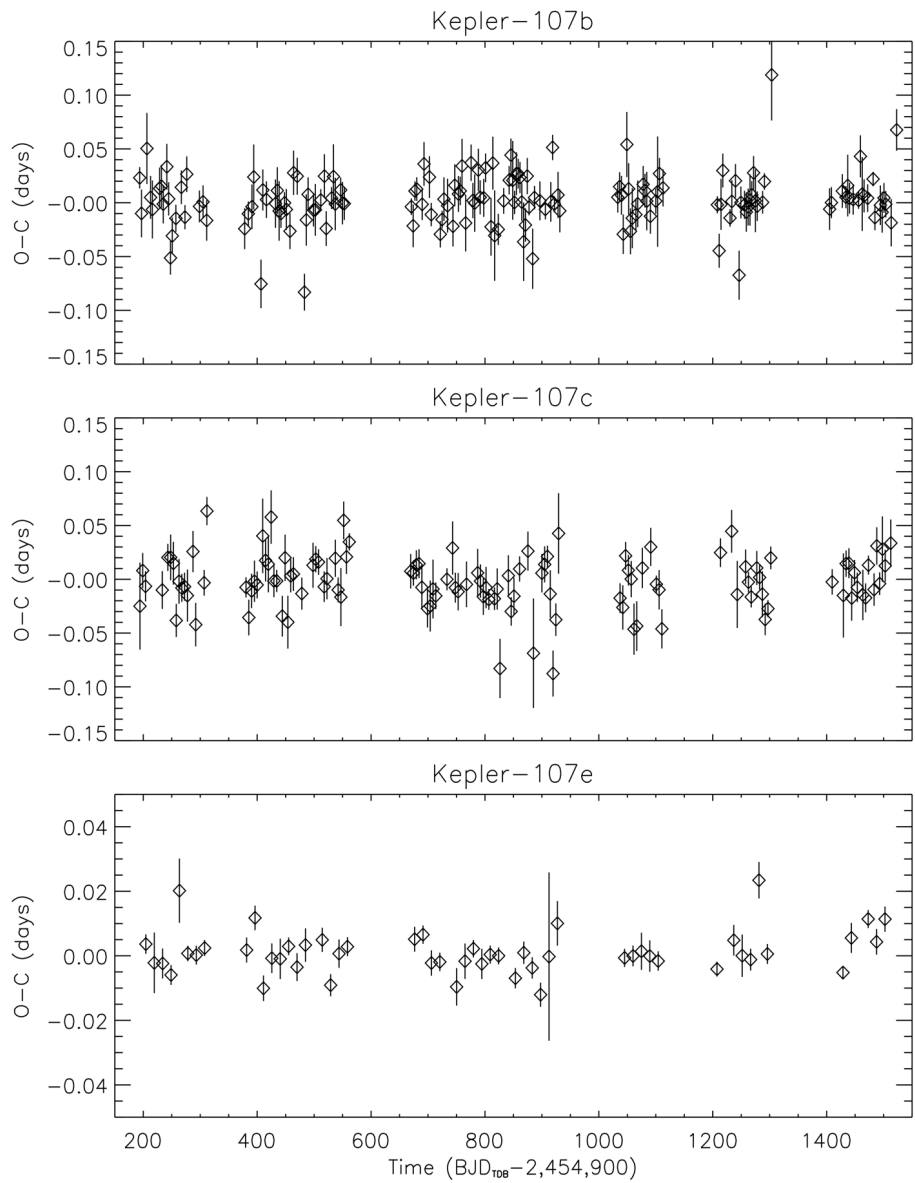

Supplementary Figure 8: **Transit Timing Variations.** The panels show, from top to bottom, the variations of the mid-transit epochs of Kepler-107b, c, and e. Those of planet d could not be computed because of the very low signal-to-noise ratio of its individual transits. At the achieved precision, no significant trends are seen.

Supplementary Table 1: **Priors on the free parameters of the two radial-velocity models.**

| Parameter | 4 Keplerians + 1 Sinusoid | 4 Keplerians + Gaussian Process regression |
|---|---|---|
| $P_b$ (d) | $\mathcal{N}(3.1800218, 2.9\times10^{-6})$ | $\mathcal{N}(3.1800218, 2.9\times10^{-6})$ |
| $T_{c,b}$ - 2,450,000 (BJD$_{TDB}$) | $\mathcal{N}(5701.08414, 3.7\times10^{-4})$ | $\mathcal{N}(5701.08414, 3.7\times10^{-4})$ |
| $K_b$ (m/s) | $\mathcal{U}[0, +\infty]$ | $\mathcal{U}[0, 10]$ |
| $P_c$ (d) | $\mathcal{N}(4.901452, 1.0\times10^{-5})$ | $\mathcal{N}(4.901452, 1.0\times10^{-5})$ |
| $T_{c,c}$ - 2,450,000 (BJD$_{TDB}$) | $\mathcal{N}(5697.01829, 7.9\times10^{-4})$ | $\mathcal{N}(5697.01829, 7.9\times10^{-4})$ |
| $K_c$ (m/s) | $\mathcal{U}[0, +\infty]$ | $\mathcal{U}[0, 10]$ |
| $P_d$ (d) | $\mathcal{N}(7.95839, 1.2\times10^{-4})$ | $\mathcal{N}(7.95839, 1.2\times10^{-4})$ |
| $T_{c,d}$ - 2,450,000 (BJD$_{TDB}$) | $\mathcal{N}(5702.9547 \pm 0.0060)$ | $\mathcal{N}(5702.9547 \pm 0.0060)$ |
| $K_d$ (m/s) | $\mathcal{U}[0, +\infty]$ | $\mathcal{U}[0, 10]$ |
| $P_e$ (d) | $\mathcal{N}(14.749143, 1.9\times10^{-5})$ | $\mathcal{N}(14.749143, 1.9\times10^{-5})$ |
| $T_{c,e}$ - 2,450,000 (BJD$_{TDB}$) | $\mathcal{N}(5694.48550, 4.6\times10^{-4})$ | $\mathcal{N}(5694.48550, 4.6\times10^{-4})$ |
| $K_e$ (m/s) | $\mathcal{U}[0, +\infty]$ | $\mathcal{U}[0, 10]$ |
| $P_{sin}$ (d) | $\mathcal{U}(13.6, 14.6)$ | |
| $T_{0,sin}$ - 2,450,000 (BJD$_{TDB}$) | $\mathcal{U}[0, +\infty]$ | |
| $K_{sin}$ (m/s) | $\mathcal{U}[0, +\infty]$ | |
| $h$ (m/s) | | $\mathcal{U}[0, 10]$ |
| $\theta$ (d) | | $\mathcal{U}[12, 16]$ |
| $w$ (d) | | $\mathcal{U}[0, 1]$ |
| $\lambda$ (d) | | $\mathcal{U}[0, 1500]$ |
| RV jitter (m/s) | $\mathcal{U}[0, +\infty]$ | $\mathcal{U}[0, 10]$ |
| $V_r$ (km/s) | $\mathcal{U}[-\infty, +\infty]$ | $\mathcal{U}[5500, 5750]$ |

Notes:

$\mathcal{N}(\mu, \sigma)$: normal distribution with mean μ and standard deviation σ

$\mathcal{U}[a, b]$: uniform distribution between a and b values

Supplementary Table 2: **Results of the radial-velocity modelling.** Error bars and upper limits refer to 1σ uncertainties.

| Parameter | 4 Keplerians + 1 Sinusoid | 4 Keplerians + Gaussian Process regression |
|---|---|---|
| $K_b$ (m/s) | 1.32 ± 0.57 | 1.39 ± 0.50 |
| $K_c$ (m/s) | 3.06 ± 0.57 | 3.15 ± 0.50 |
| $K_d$ (m/s) | < 1.1 | < 0.9 |
| $K_e$ (m/s) | 1.95 ± 0.82 | 2.56 ± 0.71 |
| $M_{p,b}(M_\oplus)$ | 3.51 ± 1.52 | 3.67 ± 1.32 |
| $M_{p,c}(M_\oplus)$ | 9.39 ± 1.77 | 9.64 ± 1.49 |
| $M_{p,d}(M_\oplus)$ | < 3.8 | < 3.1 |
| $M_{p,e}(M_\oplus)$ | 8.6 ± 3.6 | 11.3 ± 3.1 |
| $P_{sin}$ (d) | $14.10^{+0.25}_{-0.30}$ | |
| $T_{0,sin}$ - 2,450,000 (BJD$_{TDB}$) | 6827.7 ± 1.3 | |
| $K_{sin}$ (m/s) | 2.41 ± 0.70 | |
| $h$ (m/s) | | $1.77^{+1.33}_{-0.93}$ |
| $\theta$ (d) | | $14.27^{+0.36}_{-0.60}$ |
| $w$ (d) | | 0.61 ± 0.26 |
| $\lambda$ (d) | | $867^{+396}_{-492}$ |
| RV jitter (m/s) | < 0.8 | < 0.9 |
| $V_r$ (m/s) | 5644.23 ± 0.45 | $5644.30^{+1.14}_{-0.96}$ |

Supplementary Table 3: **Stellar photospheric abundances relative to the Sun.**

| Element [X/H] | Abundance [dex] | Number of lines |
|:---:|:---:|:---:|
| NaI | 0.400 ± 0.070 | 3 |
| MgI | 0.358 ± 0.049 | 3 |
| AlI | 0.389 ± 0.014 | 2 |
| SiI | 0.361 ± 0.039 | 14 |
| CaI | 0.313 ± 0.031 | 12 |
| ScI | 0.441 ± 0.007 | 3 |
| ScII | 0.425 ± 0.093 | 6 |
| TiI | 0.350 ± 0.062 | 24 |
| TiII | 0.359 ± 0.051 | 6 |
| MnI | 0.437 ± 0.063 | 5 |
| CrI | 0.352 ± 0.047 | 21 |
| CrII | 0.253 ± 0.077 | 3 |
| VI | 0.426 ± 0.010 | 6 |
| CoI | 0.452 ± 0.028 | 8 |
| NiI | 0.406 ± 0.025 | 40 |

Supplementary Table 4: **Measurements of radial velocity and activity indexes.**

| Time (BJD$_{UTC}$-2,450,000) | RV (m/s) | σRV (m/s) | FWHM (m/s) | BIS (m/s) | log(R'$_{HK}$) | σlog(R'$_{HK}$) |
|---|---|---|---|---|---|---|
| 6829.641319 | 5640.11 | 3.97 | 7796.31 | 24.90 | -4.994 | 0.080 |
| 6831.546838 | 5639.56 | 3.26 | 7812.65 | 8.89 | -5.058 | 0.070 |
| 6832.592840 | 5649.14 | 3.77 | 7803.99 | 6.29 | -5.046 | 0.086 |
| 6845.598973 | 5640.18 | 3.01 | 7820.68 | 13.08 | -5.084 | 0.066 |
| 6846.641739 | 5637.14 | 3.74 | 7819.36 | 7.02 | -5.037 | 0.076 |
| 6847.633230 | 5643.25 | 6.30 | 7823.76 | -5.93 | -5.336 | 0.320 |
| 6849.586407 | 5642.12 | 5.67 | 7824.36 | 3.06 | -5.085 | 0.154 |
| 6850.637493 | 5639.78 | 5.61 | 7806.19 | 18.41 | -5.029 | 0.132 |
| 6851.632235 | 5646.18 | 3.42 | 7809.08 | -3.65 | -5.078 | 0.078 |
| 6852.633376* | 5654.00 | 5.06 | 7797.71 | 5.36 | -5.036 | 0.128 |
| 6853.635316* | 5652.50 | 4.14 | 7812.85 | 12.61 | -4.955 | 0.078 |
| 6862.570003 | 5646.30 | 4.13 | 7816.29 | 6.92 | -5.234 | 0.148 |
| 6863.531728 | 5647.75 | 4.48 | 7837.46 | 19.71 | -5.111 | 0.124 |
| 6864.538823 | 5647.43 | 3.19 | 7803.21 | 6.28 | -5.079 | 0.072 |
| 6866.541785 | 5642.30 | 5.84 | 7815.92 | 14.80 | -5.266 | 0.241 |
| 7180.680532* | 5643.37 | 3.72 | 7799.86 | 17.13 | -5.060 | 0.083 |
| 7181.593700 | 5650.47 | 5.40 | 7836.01 | -10.56 | -5.007 | 0.123 |
| 7182.557541 | 5646.70 | 3.31 | 7814.09 | 5.08 | -5.194 | 0.098 |
| 7183.584880 | 5635.22 | 4.96 | 7831.50 | 30.06 | -5.104 | 0.129 |
| 7185.572111 | 5646.34 | 4.81 | 7799.70 | -2.05 | -5.070 | 0.123 |
| 7186.594367 | 5640.02 | 3.18 | 7783.43 | 0.97 | -5.139 | 0.083 |
| 7188.656460 | 5635.67 | 4.57 | 7782.68 | 2.48 | -4.955 | 0.092 |
| 7189.648563 | 5638.90 | 7.85 | 7795.30 | -18.51 | -5.248 | 0.326 |
| 7190.663098 | 5647.35 | 3.43 | 7799.31 | 5.55 | -5.060 | 0.080 |
| 7191.663534 | 5642.24 | 3.32 | 7792.51 | 10.11 | -5.052 | 0.073 |
| 7192.660186 | 5642.85 | 3.47 | 7809.78 | 12.46 | -5.233 | 0.117 |
| 7193.662207 | 5640.42 | 3.09 | 7796.18 | 8.44 | -5.051 | 0.064 |
| 7195.654431 | 5651.19 | 4.00 | 7815.44 | 15.26 | -5.192 | 0.124 |
| 7221.605328 | 5642.08 | 2.91 | 7782.88 | 6.56 | -5.155 | 0.075 |
| 7222.541535 | 5642.96 | 3.84 | 7780.88 | 17.27 | -5.089 | 0.100 |
| 7223.715037 | 5634.17 | 9.55 | 7802.50 | 8.70 | -4.841 | 0.187 |
| 7227.604988 | 5650.87 | 3.75 | 7791.80 | 1.63 | -5.013 | 0.078 |
| 7228.609294 | 5648.84 | 5.40 | 7818.43 | 8.34 | -5.067 | 0.141 |
| 7230.660140 | 5644.82 | 5.03 | 7818.98 | 7.37 | -5.057 | 0.136 |
| 7254.608600 | 5657.08 | 5.64 | 7831.29 | 4.92 | -4.927 | 0.123 |
| 7256.614319 | 5640.75 | 5.04 | 7799.96 | -0.71 | -4.984 | 0.107 |
| 7257.617068 | 5644.00 | 6.71 | 7822.25 | 0.99 | -5.189 | 0.272 |

| Time (BJD$_{UTC}$-2,450,000) | RV (m/s) | σRV (m/s) | FWHM (m/s) | BIS (m/s) | log(R'$_{HK}$) | σlog(R'$_{HK}$) |
|---|---|---|---|---|---|---|

Supplementary Table 4 - continued from previous page

| Time (BJD$_{UTC}$-2,450,000) | RV (m/s) | σRV (m/s) | FWHM (m/s) | BIS (m/s) | log(R'$_{HK}$) | σlog(R'$_{HK}$) |
|---|---|---|---|---|---|---|
| 7267.552112 | 5647.20 | 3.78 | 7815.64 | 13.86 | -5.131 | 0.101 |
| 7269.515188 | 5653.99 | 3.99 | 7817.17 | -1.37 | -5.013 | 0.072 |
| 7270.498495 | 5641.88 | 3.70 | 7804.91 | 13.73 | -5.033 | 0.059 |
| 7271.501443 | 5641.57 | 3.12 | 7805.10 | 17.01 | -5.309 | 0.247 |
| 7272.540536 | 5636.45 | 5.08 | 7798.88 | 12.80 | -5.049 | 0.074 |
| 7273.516979 | 5647.84 | 3.59 | 7818.20 | -0.50 | -5.125 | 0.091 |
| 7301.493922 | 5638.15 | 3.60 | 7808.70 | 15.82 | -5.028 | 0.137 |
| 7302.495732 | 5631.49 | 5.92 | 7812.62 | 32.53 | -5.164 | 0.200 |
| 7498.703402 | 5650.04 | 6.04 | 7804.65 | 24.01 | -4.905 | 0.091 |
| 7499.715106* | 5648.31 | 5.01 | 7771.64 | 23.71 | -5.449 | 0.235 |
| 7521.650229 | 5642.85 | 4.35 | 7778.49 | 6.61 | -5.012 | 0.084 |
| 7522.698371 | 5646.94 | 3.91 | 7795.96 | 27.16 | -5.089 | 0.161 |
| 7525.687954 | 5644.80 | 5.77 | 7779.45 | 8.90 | -4.966 | 0.156 |
| 7526.688837 | 5635.56 | 6.95 | 7799.03 | 13.43 | -4.961 | 0.105 |
| 7527.660506* | 5644.00 | 5.12 | 7784.03 | 8.00 | -5.262 | 0.182 |
| 7528.680232* | 5647.21 | 4.63 | 7789.27 | 23.25 | -5.118 | 0.157 |
| 7529.691289* | 5649.94 | 5.25 | 7792.05 | 33.71 | -4.968 | 0.134 |
| 7530.694093* | 5641.97 | 6.03 | 7822.29 | 25.68 | -5.242 | 0.202 |
| 7557.662814* | 5649.88 | 5.14 | 7802.55 | -7.13 | -4.955 | 0.146 |
| 7558.633191* | 5641.31 | 6.84 | 7794.55 | 41.52 | -5.019 | 0.157 |
| 7559.616508* | 5650.48 | 7.19 | 7802.48 | 7.84 | -5.103 | 0.245 |
| 7565.585627 | 5644.46 | 8.22 | 7799.69 | 9.43 | -5.093 | 0.071 |
| 7566.606557 | 5644.40 | 3.11 | 7785.22 | 1.69 | -5.085 | 0.131 |
| 7573.530542 | 5647.73 | 4.91 | 7805.09 | 7.86 | -5.075 | 0.111 |
| 7573.551734 | 5640.74 | 4.42 | 7774.72 | 23.15 | -4.964 | 0.064 |
| 7574.521167 | 5638.21 | 3.49 | 7803.44 | 12.80 | -5.061 | 0.108 |
| 7574.541491 | 5633.35 | 4.56 | 7787.33 | 14.86 | -5.080 | 0.104 |
| 7576.512913 | 5645.31 | 4.33 | 7784.56 | 9.36 | -5.014 | 0.072 |
| 7576.533388 | 5645.57 | 3.77 | 7793.50 | 9.41 | -6.314 | 3.430 |
| 7602.413100 | 5644.28 | 7.23 | 7787.16 | 9.54 | -5.147 | 0.166 |
| 7602.433285 | 5639.18 | 5.34 | 7809.63 | -3.55 | -5.290 | 0.140 |
| 7614.424583* | 5641.24 | 3.79 | 7795.82 | 7.68 | -5.100 | 0.088 |
| 7614.447303* | 5646.04 | 3.67 | 7811.11 | -12.32 | -5.032 | 0.116 |
| 7617.419103* | 5647.99 | 4.95 | 7785.28 | -20.50 | -4.968 | 0.107 |
| 7617.441696* | 5647.10 | 5.20 | 7770.38 | -36.37 | -5.158 | 0.170 |
| 7618.415801* | 5641.99 | 5.57 | 7775.99 | -64.62 | -5.148 | 0.244 |
| 7618.437109* | 5637.68 | 7.67 | 7814.93 | -16.78 | -5.062 | 0.076 |
| 7619.391064* | 5635.79 | 3.50 | 7804.13 | -17.10 | -5.113 | 0.085 |

Supplementary Table 4 - continued from previous page

| Time (BJD$_{UTC}$-2,450,000) | RV (m/s) | σRV (m/s) | FWHM (m/s) | BIS (m/s) | log(R'$_{HK}$) | σlog(R'$_{HK}$) |
|---|---|---|---|---|---|---|
| 7619.413876* | 5640.07 | 3.47 | 7788.45 | -0.28 | -5.117 | 0.109 |
| 7651.362036* | 5652.12 | 4.11 | 7811.95 | 7.79 | -4.985 | 0.070 |
| 7651.382707* | 5649.98 | 3.65 | 7783.03 | 1.69 | -5.246 | 0.120 |
| 7652.360994 | 5649.98 | 3.59 | 7808.77 | 6.72 | -4.996 | 0.061 |
| 7652.382047 | 5650.90 | 3.40 | 7794.89 | 9.01 | -5.092 | 0.184 |
| 7653.365426 | 5640.99 | 6.26 | 7803.50 | 12.56 | -4.977 | 0.151 |
| 7653.386397 | 5649.11 | 6.68 | 7796.38 | 16.06 | -5.111 | 0.120 |
| 7654.364440 | 5644.77 | 4.48 | 7791.02 | -2.01 | -5.321 | 0.226 |
| 7654.385528 | 5646.10 | 4.99 | 7791.52 | -3.24 | -5.124 | 0.178 |
| 7655.403210 | 5646.98 | 5.81 | 7786.50 | 16.71 | -5.268 | 0.339 |
| 7655.425003 | 5642.61 | 7.34 | 7795.17 | 24.01 | -5.045 | 0.081 |
| 7656.357283 | 5650.08 | 3.71 | 7767.62 | 4.63 | -5.208 | 0.131 |
| 7656.378463 | 5649.12 | 3.92 | 7808.16 | -30.68 | -4.924 | 0.147 |
| 7658.395272 | 5634.23 | 7.73 | 7794.18 | 14.02 | -5.079 | 0.217 |
| 7658.415573 | 5632.06 | 7.95 | 7775.04 | -3.57 | -5.136 | 0.141 |
| 7659.433207 | 5636.31 | 4.74 | 7798.11 | -5.26 | -5.044 | 0.122 |
| 7659.454318 | 5632.62 | 5.13 | 7772.62 | -4.62 | -5.186 | 0.090 |
| 7661.374576 | 5642.01 | 3.19 | 7790.73 | 10.60 | -5.039 | 0.065 |
| 7661.395478 | 5640.21 | 3.19 | 7790.11 | 5.15 | -5.176 | 0.089 |
| 7669.359450 | 5645.77 | 3.27 | 7793.04 | -0.61 | -5.400 | 0.160 |
| 7669.379947 | 5644.30 | 3.44 | 7767.81 | -0.68 | -5.190 | 0.197 |
| 7670.382276 | 5645.13 | 5.63 | 7807.98 | 9.98 | -5.114 | 0.150 |
| 7670.403398 | 5635.47 | 5.21 | 7769.91 | 5.61 | -4.935 | 0.069 |
| 7671.353842 | 5647.08 | 3.87 | 7778.44 | 5.99 | -5.163 | 0.094 |
| 7671.374605 | 5649.73 | 3.38 | 7795.49 | 12.10 | -5.121 | 0.090 |
| 7672.357559* | 5648.41 | 3.44 | 7760.09 | 3.41 | -4.991 | 0.071 |
| 7672.378241 | 5643.34 | 3.53 | 7771.20 | 0.38 | -5.073 | 0.089 |
| 7673.359389 | 5643.28 | 4.03 | 7797.50 | 8.51 | -5.264 | 0.133 |
| 7673.380719 | 5641.43 | 3.91 | 7802.90 | 21.48 | -5.052 | 0.090 |
| 7699.345433 | 5649.22 | 4.34 | 7750.61 | -15.18 | -5.018 | 0.143 |
| 7706.370203 | 5652.20 | 6.33 | 7790.39 | 10.95 | -5.000 | 0.130 |
| 7721.345665 | 5641.33 | 5.95 | 7780.78 | 10.85 | -5.062 | 0.099 |
| 7727.324829 | 5640.16 | 3.99 | 7813.98 | 12.55 | -4.976 | 0.123 |
| 7728.330556 | 5646.56 | 5.57 | 7793.96 | -4.52 | -4.949 | 0.111 |
| 7729.315081 | 5636.67 | 5.75 | 7820.37 | -0.95 | -5.113 | 0.175 |
| 7861.716662 | 5650.46 | 6.27 | 7803.22 | -14.40 | -5.088 | 0.130 |
| 7863.730901 | 5643.20 | 5.73 | 7814.61 | 6.50 | -5.075 | 0.082 |
| 7864.667902 | 5632.46 | 3.98 | 7807.42 | 19.87 | -5.195 | 0.272 |

Supplementary Table 4 - continued from previous page

| Time (BJD$_{UTC}$-2,450,000) | RV (m/s) | σRV (m/s) | FWHM (m/s) | BIS (m/s) | log(R'$_{HK}$) | σlog(R'$_{HK}$) |
|---|---|---|---|---|---|---|
| 7865.685104 | 5631.66 | 7.97 | 7856.53 | 6.00 | -4.777 | 0.160 |

Notes:
*: observations corrected for moonlight contamination.